\documentclass{aa}  

\usepackage{epsfig}	
\usepackage{graphicx,color}	
\usepackage{amssymb}		
\usepackage{url}		
\usepackage{amsmath}		
\usepackage{rotating}			
\usepackage{float}			
\usepackage{textcomp}		
\usepackage{epstopdf}
\usepackage{dcolumn}
\usepackage{times}
\usepackage{tabularx}
\usepackage[english]{babel}
\usepackage{hyperref}
\hypersetup{
    colorlinks,
    citecolor=blue,
    filecolor=blue,
    linkcolor=blue,
    urlcolor=blue,
    menucolor=black}

\begin{document}


\title{On the size distribution of spots within sunspot groups}
   \author{Sudip Mandal
          \inst{1}
          \and
          Natalie A. Krivova\inst{1}
          \and
          Robert Cameron\inst{1}
          \and
          Sami K. Solanki\inst{1,2}
          }

   \institute{Max Planck Institute for Solar System Research, Justus-von-Liebig-Weg 3, 37077, G{\"o}ttingen, Germany \\
              \email{smandal.solar@gmail.com}, ORCID: 0000-0002-7762-5629
         \and
             School of Space Research, Kyung Hee University, Yongin, Gyeonggi 446-701, Republic of Korea
             }

\abstract{Size distribution of sunspots provides key information about the generation and emergence processes of the solar magnetic field. Previous studies on the size distribution have primarily focused on either the whole group or individual spot areas. In this paper, we investigate the organization of spot areas within sunspot groups.
In particular, we analyze the ratio, $\rm{R}$, of the area of the biggest spot ($\rm{A_{big\_spot}}$) inside a group, to the total area of that group ($\rm{A_{group}}$). We use sunspot observations from Kislovodsk, Pulkovo and Debrecen observatories, together covering solar cycles 17 to 24. We find that at the  time  when  the  group  area reaches its maximum, the single biggest spot in a group typically occupies about 60\% of the group area. For half of all groups, $\rm R$ lies in the range between roughly 50\% and 70\%. We also find R to change with the group area, $\rm{A_{group}}$, such that $\rm{R}$ reaches a maximum of about 0.65 for groups with $\rm{A_{group}}\approx 200\mu$Hem and then remains at about 0.6 for lager groups. 
Our findings imply a scale invariant emergence pattern, providing an observational constraint on the emergence process. Furthermore,
extrapolation of our results to larger sunspot groups may have a bearing on the giant unresolved starspot features found in doppler images of highly active sun-like stars. Our results suggest that such giant features are composed of multiple spots, with the largest spot occupying roughly 55--75\% of the total group area (i.e. of the area of the giant starspots seen in Doppler images).}

   \titlerunning{On the size distribution of spots within sunspot groups}
   \authorrunning{Mandal et al.}
   \maketitle

\section{Introduction}

Magnetic field of the Sun, the driving force of its activity and variability, is generated in its interior by a dynamo action and emerges at the surface as bipolar regions. The larger of these regions typically host sunspots which are dark photospheric features that trace out the locations of highly concentrated magnetic field on the solar surface \citep{2003A&ARv..11..153S}. Direct measurements of the solar magnetic fields are only available for the last few decades, and hence, sunspot area records, which have been observed more or less regularly over more than a century, act as a proxy of the surface magnetism and provide an indirect way of understanding the long-term behaviour of the magnetic activity and variability. 

Sunspots usually emerge in groups, and within a typical group, spots are arranged into two sub-groups, leading (closer to the solar equator) and following. Spots in the leading sub-group are bigger and more coherent than following one
\citep{1979suns.book.....B,1981phss.conf....7M}. Although the reason behind such a configuration is not yet understood, it is believed to be related to the flux emergence process where various forces (e.g. Coriolis force, convective motions, sub-surface shears etc.) act on a rising flux tube and determine its surface  morphology \citep{2011LRSP....8....4B,2015LRSP...12....1V}.

One of the aspects providing clues to understanding the processes of solar magnetic field generation and emergence, is the size distribution of spots. Studies of the size distribution of both the sunspot groups and the individual spots have been carried out in the past by various authors \citep{1980BAICz..31..224K,1988ApJ...327..451B,2005A&A...443.1061B,2012ApJ...758L..20N,2014SoPh..289.1143T,2015ApJ...800...48M}.
Using the group area data from the Royal Greenwich Observatory (RGO), \citet{1988ApJ...327..451B} and \citet{2005A&A...443.1061B} found that the size distribution, over its most range, could be fairly well described by a log-normal function.
Physically, a log-normal size distribution suggests that all spots are generated by a global dynamo action, with smaller spots being the fragmented products of the larger flux concentrations \citep{1941DoSSR..30..301K}. Later, \citet{2009ApJ...698...75P} and \citet{2011SoPh..269...13T} analysed various observations of the magnetic field emergence and concluded that they could all be described by a single power-law distribution, covering seven orders of magnitude in flux. However, using individual spot area measurements from the Kislovodsk Mountain Station and group areas from RGO, Solar Observing Optical Network (SOON), Pulkovo, and Kislovodsk, \citet{2012ApJ...758L..20N} and \citet{2015ApJ...800...48M}, respectively, showed that the overall size distribution was better characterized by two separate distribution functions for `small' and `big' groups or spots. They argued that such a bi-modal size distribution could possibly imply a more complex process of spot formation, e.g. driven by small-scale and global dynamos in parallel \citep{2016ApJ...833...94N}. A similar conclusion was also reached by \citet{2014SoPh..289.1143T} who analysed the magnetic flux and sunspot area records from Helioseismic and Magnetic Imager (HMI).
However, results from recent high-resolution observations and  numerical simulations suggest that small-scale dynamo is mainly responsible for the generation of the (non-spot) weak fields \citep{2010ApJ...714.1606P,2012A&A...547A..93S,2014ApJ...789..132R,2016ApJ...816...28K}.

Thus, understanding the spot size distribution calls for further investigations. Furthermore, understanding the spot area distribution on the Sun can help us to understand the poorly known area distribution of starspots on other Sun-like stars \citep[e.g.,][]{2004MNRAS.348..307S}. In this paper, we re-address the question of the size distribution of sunspots. However, instead of analysing the overall spot or group size distribution as done in previous studies, we here focus on the distribution of spot areas within individual groups.
 We describe the data and the approach used in this study in Sect.~2 and the results in Sect.~3, while Sect.~4 summarises our conclusions.

\section{Data and Methods}

\begin{figure*}[!htb]
\centering
\includegraphics[width=0.90\textwidth,clip,trim=0cm 5cm 0cm 0cm]{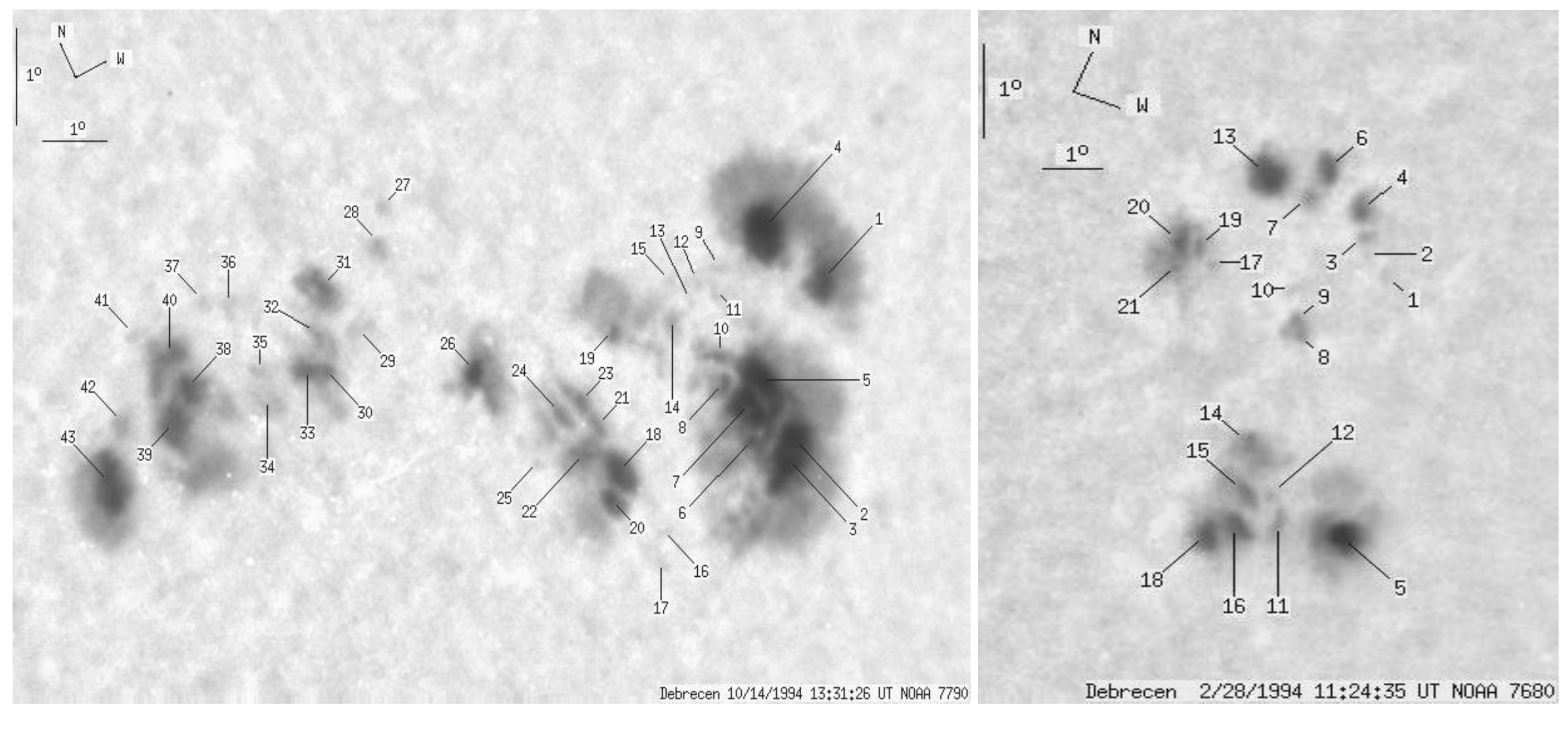}
\label{new_fig}
\caption[]{Images of two sunspot groups (NOAA 7790 and NOAA 7680) from Debrecen observatory\footnotemark. In each image, individual umbrae are marked by unique numbers.}
\end{figure*}

We use sunspot area catalogues from three different sources, namely, Debrecen Observatory\footnote{\label{DPD1}\url{http://fenyi.solarobs.csfk.mta.hu/en/databases/DPD/}} \citep{2016SoPh..291.3081B,2017MNRAS.465.1259G}, Kislovodsk Mountain station\footnote{\url{http://158.250.29.123:8000/web/Soln_Dann/}} \citep{2007SoSyR..41...81N} and Pulkovo observatory\footnote{\url{http://www.gaoran.ru/database/csa/groups_e.html}} \citep{1955Obs....75...28M}. Kislovodsk (1954--2019) and Pulkovo (1932--1991) catalogues provide daily records of the total area of every individual group, $\rm{A_{group}}$ (including umbrae and penumrae), the total area of the biggest spot, $\rm{A_{big\_spot}}$ (including umbra and penumra), as well as the number, $\rm{N}$, of individual umbrae in each group. Since some spots might include multiple umbrae (i.e. multiple umbrae embedded within the same penumbra, e.g., see Fig.~\ref{new_fig}), for some groups $\rm{N}$ can be higher than the number of the individual spots in the group.
 The Debrecen record is the official continuation of the RGO programme after it was ceased in 1976 \citep{2016SoPh..291.3081B}.
This catalogue\footnote{Further details are available in \url{http://fenyi.solarobs.csfk.mta.hu/ftp/pub/DPD/DPDformat.txt}} lists daily measurements of  $\rm{A_{group}}$ and $\rm{N}$ values. Each spot, in this catalogue, carries a unique group label using which we calculate $\rm{A_{big\_spot}}$ from Debrecen as well.

\footnotetext{These images are available at \url{http://fenyi.solarobs.csfk.mta.hu/en/databases/DPD/}}

 Since area measurements from different observatories show systematic differences \citep{1997SoPh..173..427F,2001MNRAS.323..223B,2009JGRA..114.7104B}, a cross-calibration is necessary before using them for any analysis. Various past studies have found that both the daily and the individual area measurements from the catalogues considered here (Deberecen, Kislovodsk and Pulkovo) are very similar to each other, and also to the measurements from RGO, with mutual cross-calibration factors being close to unity \citep{2007SoSyR..41...81N,2009JGRA..114.7104B,2013MNRAS.434.1713B,2015ApJ...800...48M,2020A&A...640A..78M}. For the purpose of this work, we use the latest calibration by \citet{2020A&A...640A..78M} who compared these datasets statistically over the overlapping periods to generate a consistent and homogeneous area record.

 To understand the area distribution of spots within individual groups, we focus on the biggest spot in the group and derive the ratio, $\rm{R}$, between the area of the biggest spot and the total group area: 
 \begin{equation}
     \rm{R=A_{big\_spot} / A_{group}}.
     \label{eqn1}
 \end{equation}
According to this definition, the maximum possible value of $\rm{R}$ is 1, for groups with only one listed spot. Let us remind here that, throughout this paper, the term `spot' or `sunspot' refers to a structure that has a solitary umbra or multiple umbrae embedded within a single penumbra.  All the areas used in this study are corrected for the foreshortening effect.

\section{Results}
\subsection{Instantaneous distributions}

\begin{figure*}[!htb]
\centering
\includegraphics[width=0.80\textwidth]{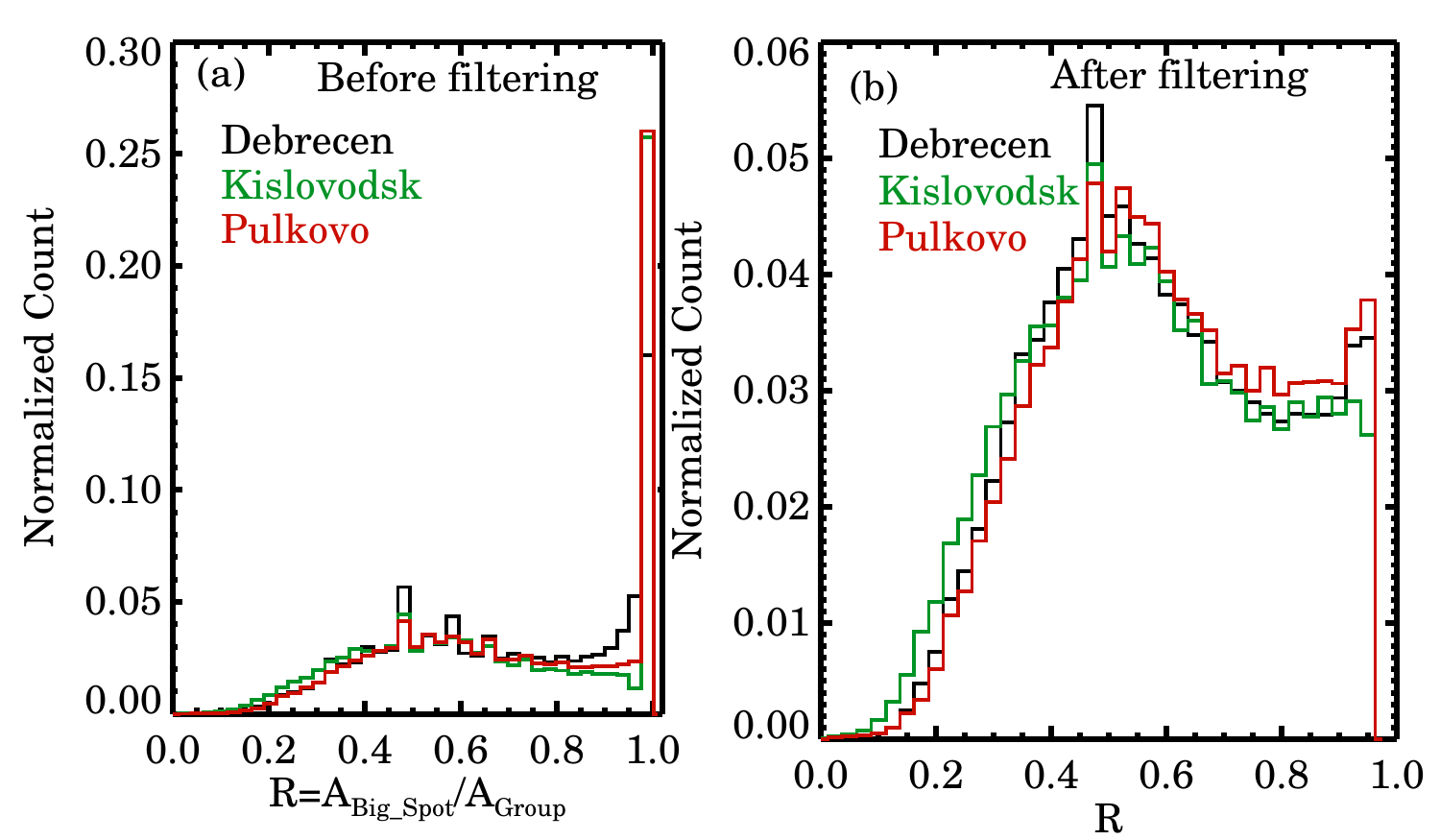}
\caption{(a) The normalised distribution of $\rm{R}$ from Debrecen (black), Kislovodsk (green) and Pulkovo (red), when considering all daily individual groups, i.e. irrespective of their evolutionary phase. (b) Same as (a), but after removing groups with $\rm{N}$=1 and $\rm{A_{group}}\leq$30~$\mu$Hem. }
\label{fig_non_tracking}
\end{figure*}

  We start by considering all data on every given day, i.e. irrespective of the group's evolutionary phase on that day. $\rm{R}$ values are then calculated for all individual group entries. Figure~\ref{fig_non_tracking}a shows the derived distributions of $\rm{R}$ from Debrecen (black histogram), Kislovodsk (green) and Pulkovo (red) data. All three distributions display some common features, e.g., a sharp primary peak at R=1 and a flatter secondary peak around $\rm{R}=0.5$. The peak at $\rm{R}$=1 comes from the single-spot groups for which $\rm{A_{big\_spot}}$= $\rm{A_{group}}$. Such groups make approximately 19\% of all the listed entries in our catalogues. Since we are interested in the distribution of spots within a group, we remove these single spot groups from our sample. In addition to that, we also reject small groups, with area less than 30~$\mu$Hem, as they are likely to be associated with larger measurement uncertainties  \citep[see][]{2015ApJ...800...48M}.
  We note, however, that a change of this threshold to different values within the range 30--80~$\mu$Hem led to the results only marginally different from those presented below, while reducing the statistics.

  The final distributions, after implementing these two restrictions, are shown in Fig.~\ref{fig_non_tracking}b. Two distinct features are evident in all three distributions. First, the distributions are clearly asymmetric, being skewed towards higher ratios (skewness=0.03) and their mode values lying around $\rm{R}\approxeq$0.5. Secondly, although the peak at $\rm{R}$=1 has now disappeared, a weak secondary peak is still seen at R$\approxeq$0.97, which is pronounced in Debrecen and Pulkovo data and is significantly weaker in Kislovodsk data.
 We will return to this peak in Sect.~3.2.

The shape of these histograms might possibly be affected by the complex evolutionary scenarios of sunspot groups, such as fragmentation or coalescence \citep{1997ApJ...487..424S}. Furthermore, while spots emerge relatively fast, their decay times are much longer \citep{1992SoPh..137...51H}. Thus, as has been pointed out in the past,  the results derived from a snapshot distribution (i.e. considering all spots independently of their evolution) can potentially be dominated by the longer decay phase of sunspot groups \citep{2005A&A...443.1061B}. To examine the influence of these effects on the obtained distributions, we track every group in our catalogues and re-derive the $\rm{R}$ distributions such that each group is considered only once.

\subsection{Tracking individual sunspot groups}

\begin{figure*}[!htbp]
\centering
\includegraphics[width=0.98\textwidth]{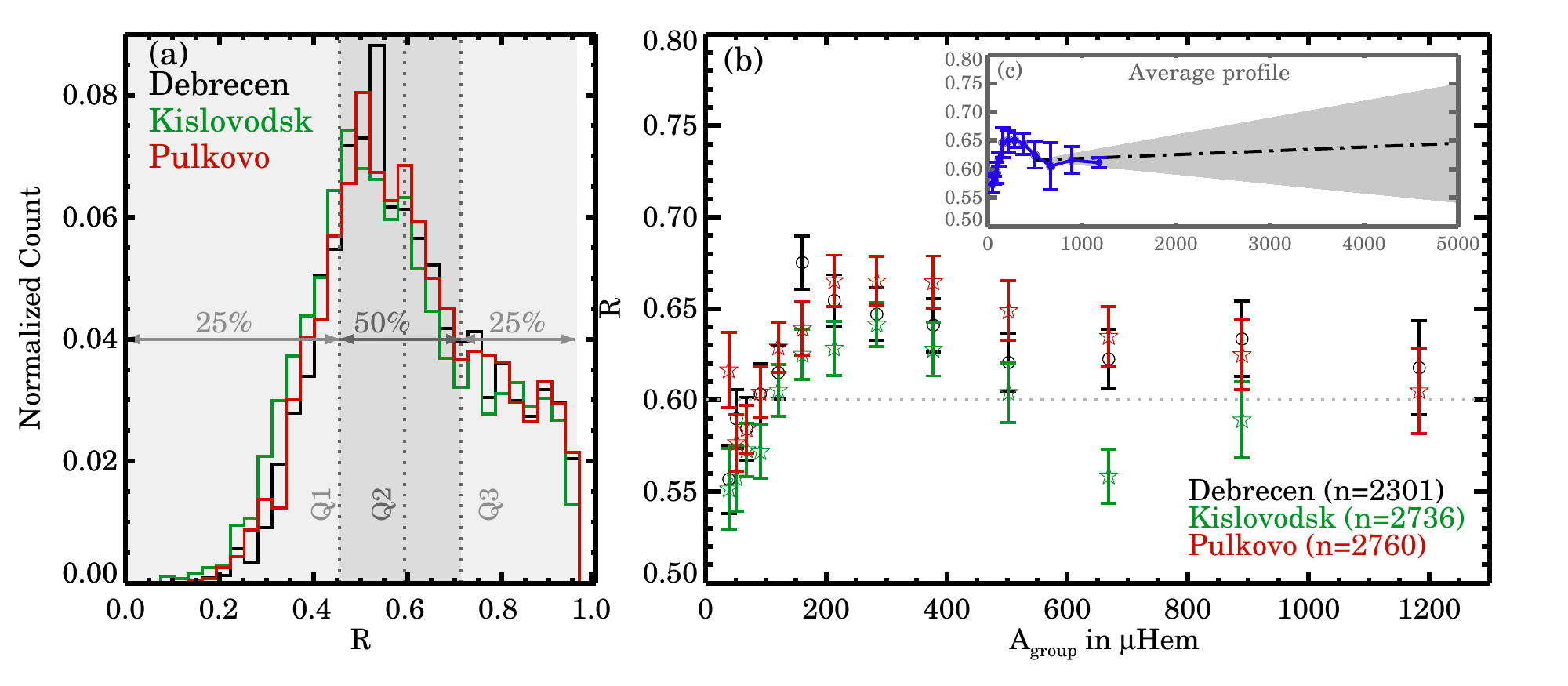}
\caption{ (a) Normalised distribution of $\rm{R}$ calculated at the time of maximum $\rm{A_{group}}$. (b) $\rm{R}$ as a function of $\rm{A_{group}}$. Error bars represent the standard deviations calculated for each bin. (c) Same as (b) but averaged over the three individual datasets. Dashed line is the linear fit to the data between 500--1200~$\mu$Hem, whereas the grey shading highlights the 1-$\sigma$ uncertainty of the fit.}
\label{fig_with_tracking}
\end{figure*}

We first visually inspected a subset of Debrecen images\footnote{Available here \url{ftp://ftp.ngdc.noaa.gov/STP/SOLAR_DATA/SUNSPOT_REGIONS/Debrecen/Debrecen_Photoheliographic_Data/images/}} from the period 1986--1999, which trace individual groups.
These images are extractions of the full-disc photographs and within each of these snapshots, spots forming a group are uniquely numbered.
 These images allowed us to trace out 40 individual groups (chosen randomly).
 By following these groups from their first emergence to their disappearance, we have noticed that, during the initial growth or final decay phase, the following spots are significantly smaller than the leading spots.
 In other words, as has also been noticed in earlier studies, leading spots form faster than the following ones \citep{1981phss.conf....7M,2014ApJ...785...90R}, and decay more slowly that following spots \citep{1963BAICz..14...91B}. This leads to $\rm{R}$ values very close to (though not exactly) 1 during these periods. This is, in fact, the reason behind the weak secondary peak at R$\sim$0.97 which we see in the instantaneous distributions (Figure~\ref{fig_non_tracking}b).

 Therefore, we now use the information from all the  catalogues considered here to track sunspot groups over their lifetime. To minimise the errors introduced by the foreshortening corrections, only groups within central meridian distances of $\pm$65\textdegree are considered here. Following the snapshot analysis described in the previous section, we leave out single-spot ($\rm{N}$=1) groups, and groups with area less than 30~$\mu$Hem.
Our aim here is to consider every group only once over its entire lifetime.
Usually, such studies consider the time when a group area ($\rm{A_{group}}$) reaches its maximum value. However, in our case, the second parameter entering $\rm{R}$, the size of the biggest spot $\rm{A_{big\_spot}}$, also evolves with time 
and $\rm{A_{big\_spot}}$ and $\rm{A_{group}}$ do not necessarily reach their maximum at the same time.
In fact, we find that only 65\% of the total groups analyzed here, have the maximum of these two quantities on the same day.
For 20\% of all groups, $\rm{A_{big\_spot}}$ reaches its maximum after the maximum in $\rm{A_{group}}$ (termed a positive delay here) whereas 15\% of the remaining groups show a negative delay.  Both, the positive and the negative delays mainly happen in big complex groups with many spots, which require a separate study. For the purpose of this paper, we only consider groups with no delay, i.e. when the groups and the biggest spot areas reach their maximum on the same day.

The distribution of $\rm{R}$ derived for these groups is shown in Figure~\ref{fig_with_tracking}a. The maxima of the distributions in all three cases lie around $\rm{R}$= 0.5, and unlike the snapshot distributions, there are no secondary peaks. To further understand these distributions, we calculate the quartiles Q1, Q2 and Q3, which describe the sample 25, 50 and 75 percentiles, respectively  \citep{2000ipse.book.....R}; see Table~\ref{table1}.
In Figure~\ref{fig_with_tracking}a, the shaded regions highlight the quartiles computed for the distribution averaged over the three datasets. The range between Q1 and Q3 represents the interquartile range (IQR=Q3-Q1), where 50\% of the data points around the median value, Q2, lie. IQR is considered to be a measure of the spread of the distribution. For all the individual and the average distributions, IQR is only 0.26. Such a low IQR indicates that the data are clustered about the central tendency, i.e. the median value (Q2) of $\rm{R}$=0.6. Thus, in the majority of groups, a single large spot occupies about 48--74\% of the total group area.

\begin{table*}[h]
    \centering
    \begin{tabular}{lccccccc}
    \hline
        Observatory & Q1 & Q2 & Q3 & Mean & Mode  & Variance & Skewness\\
    \hline
        Debrecen   & 0.49 & 0.59 & 0.74 & 0.62 & 0.54  & 0.03 & 0.26\\
        Kislovodsk & 0.47 & 0.58 & 0.73 & 0.60 & 0.49  & 0.03 & 0.22\\
        Pulkovo    & 0.50 & 0.60 & 0.75 & 0.62 & 0.53  & 0.03 & 0.22\\
    \hline    
        Average distribution & 0.48 & 0.60 & 0.74 & 0.61 & 0.52 & 0.03 & 0.21\\
    \hline
    \end{tabular}
    \caption{ Parameters of the $\rm{R}$ distributions shown in Fig. 2a.
    }
    \label{table1}
\end{table*}

Some properties of sunspot groups, such as their rotation rates \citep{1986A&A...155...87B}, growth and decay rates \citep{1992SoPh..137...51H,2008SoPh..250..269H,2014SoPh..289..563M,2020ApJ...892..107M,2021ApJ...908..133M}, hemispheric asymmetry \citep{2016ApJ...830L..33M} etc, have been found to show a clear dependence on the group sizes. Therefore, we now investigate whether a similar behaviour also exists for $\rm{R}$.
To examine this, we bin the group areas (in bins of equal widths in $\log(A_{group})$) and calculate the median value of $\rm{R}$ in each bin. Figure~\ref{fig_with_tracking}b shows the dependence of $\rm{R}$ on $\rm{A_{group}}$ for the three observatories.
Barring the first few points below 100~$\mu$Hem and a single bin centred at 650~$\mu$Hem for Kislovodsk, $\rm{R}$ stays above 0.6 over the entire range of sunspot group sizes.
We also observe an interesting three-phase evolution in $\rm{R}$. First, as $\rm{A_{group}}$ increases from 30 to 200~$\mu$Hem, $\rm{R}$ rises rapidly from about 0.55 to 0.65. Between about $\rm{A_{group}}$ of 200 and 600~$\mu$Hem it decreases somewhat and then eventually settles at a constant value of around 0.6 beyond $\rm{A_{group}}\geq$600~$\mu$Hem.
 Such a trend can be explained by the fact that most of the small groups ($\rm{A_{group}}<$50~$\mu$Hem) have a simple bipolar structure with roughly equal distribution of areas between the two constituent spots and thus $\rm{R}\approxeq$0.5. As $\rm{A_{group}}$ increases, the compact leading spot grows rapidly \citep{1981phss.conf....7M,2014ApJ...785...90R}.
This leads to an increase in $\rm{R}$ value and explains the initial rise of R from 0.55 to 0.65 below $\rm{A_{group}}\approxeq$200~$\mu$Hem. The drop beyond that to a constant value of 0.6 is unexpected and interesting. We suggest that this aspect of the data indicates a degree of scale-invariance in the flux emergence process, i.e a large active region is just a scaled up version of a small active region, with the size of all the spots increasing in proportion. In this interpretation, the slow decrease in R seen for $\rm{A_{group}}\geq$200~$\mu$Hem is a consequence of the larger bright-points in a small active region being scaled up in flux to become small spots in larger active regions.

Our findings are also of relevance for understanding giant starspots seen on highly active (rapidly rotating) sun-like stars (see, e.g., \citealp{2005LRSP....2....8B,2009A&ARv..17..251S}). In Fig.~\ref{fig_with_tracking}c, we show an extrapolation of the $\rm{R}$ vs $\rm{A_{group}}$ relation up to $\rm{A_{group}}$=5000~$\mu$Hem. We perform this extrapolation by applying a linear fit to the averaged data (i.e averaged over the three individual datasets) over the range $\rm{A_{group}}$=500--1200~$\mu$Hem. The grey shaded region in Fig.~\ref{fig_with_tracking}c highlights the 1-$\sigma$ uncertainty associated with this fitting. We find that the largest spot within a huge group (of $\approx$5000~$\mu$Hem) would cover roughly 55--75\% of the group area. These giant, though unresolved, spot structures have already been observed on highly active (rapidly rotating) sun-like stars and thus, our result puts constrains on the properties of such starspots. We stress, however, that this is a phenomenological projection of the solar case to other stars, and other factors might play a role in determining the size of individual spots on such stars.

\begin{figure}[!htb]
\centering
\includegraphics[width=0.45\textwidth]{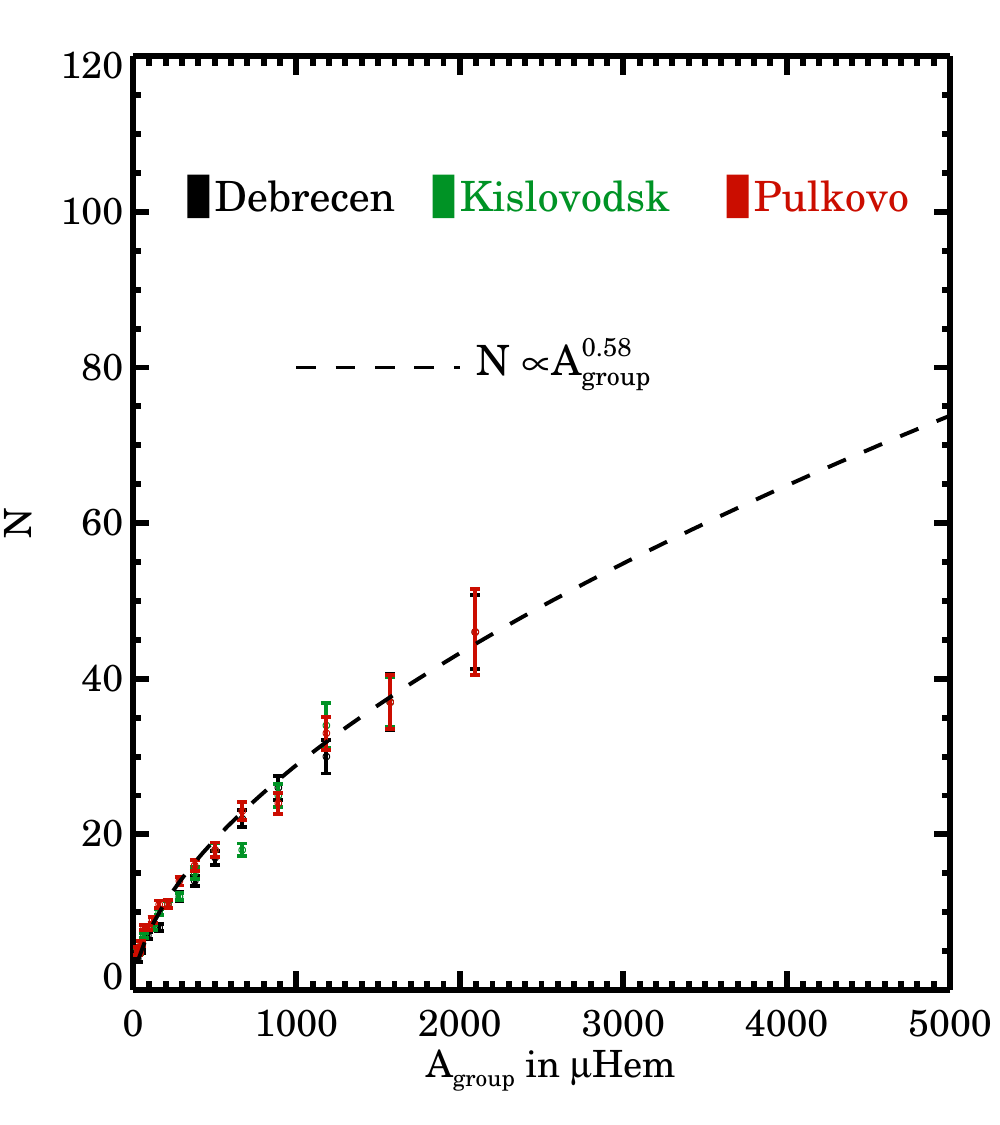}
\caption{$\rm{N}$ as a function of group area $\rm{A_{group}}$. The dash line shows the best power law fit to the data.
}
\label{fig_ratio_number}
\end{figure}

Thus far we have shown that, at maximum group growth, approximately 60\% of the total group area is occupied by a single big spot. We now analyse the spot distribution within the remaining 40\% of the group area. For this, we consider the relationship between $\rm{N}$ and $\rm{A_{group}}$. We bin the group areas $\rm{A_{group}}$ in log scale and calculate the average values of $\rm{N}$ in each bin. The result is plotted in Figure~\ref{fig_ratio_number}.
Overall, the number of individual umbrae within groups, $\rm{N}$, increases with $\rm{A_{group}}$. The rate of increase slows down with the increasing group area and the overall profile can be described with a power law function, $\rm{N}$$\propto$$\rm{A^{0.58}_{group}}$ (dashed line in Fig.~\ref{fig_ratio_number}). 
Since $\rm{N}$ is the number of umbrae within a given group while some spots have multiple penumbrae, the dependence of the number of spots within a group on its area would be somewhat flatter than $\rm{N}(\rm{A_{group}})$. Our result implies that, with the increasing group area, new flux emerges in the form of new spots (such that the number of spots in a group increases), while the primary biggest spot within the group keeps growing, too (which keeps the $\rm{R}$ constant).

\subsection{Variation of ${\rm{R}}$ with solar activity}

\begin{figure*}[!htb]
\centering
\includegraphics[width=0.99\textwidth]{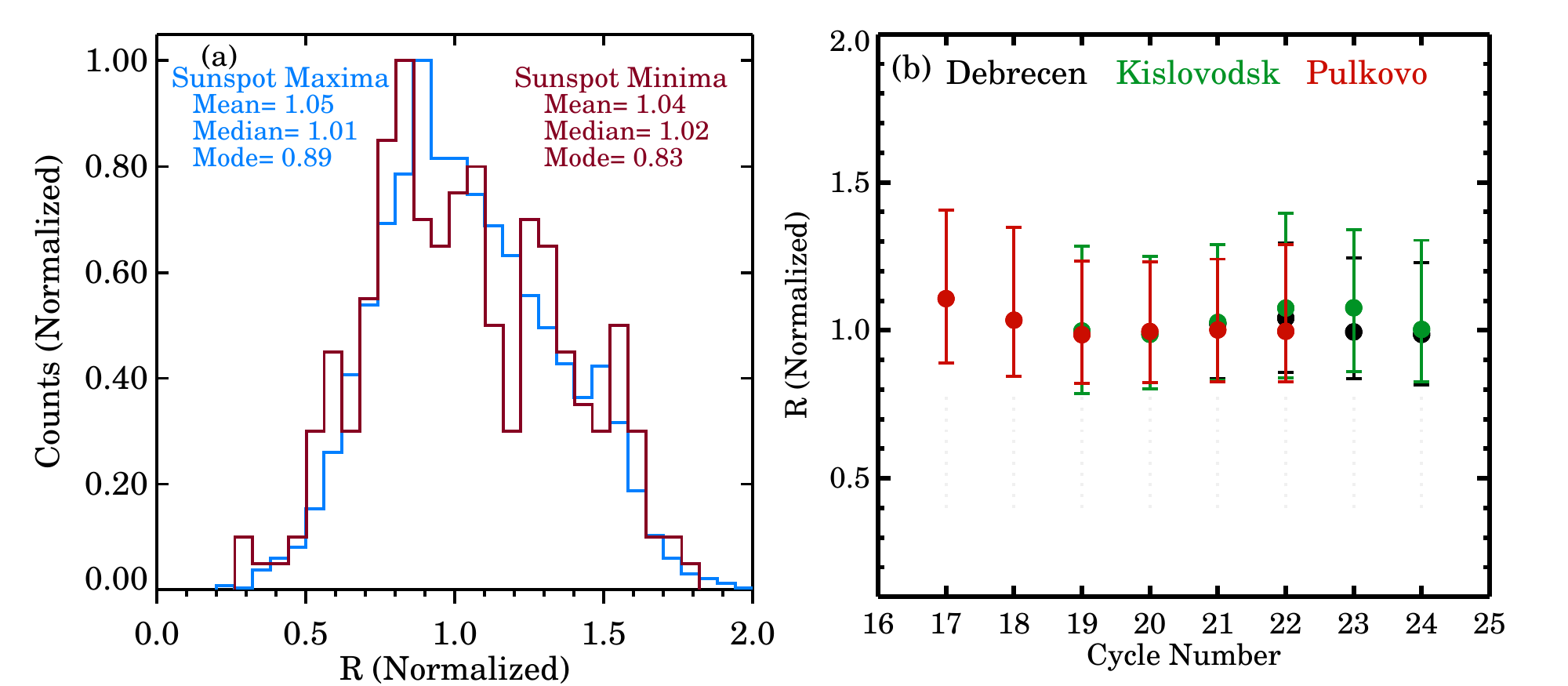}
\caption{a: The distributions of $\rm{R}$ during sunspot maxima (in blue) and minima (in red) for the merged area data. b: Cycle to cycle variation of $\rm{R}$ (median) for the three individual observatories: Debrecen (black), Kislovodsk (green) and Pulkovo (red). Upper and lower limits of error bars represent the third and first quartiles of corresponding $\rm{R}$ distributions.}  
\label{fig_cycle}
\end{figure*}

It is well known that many sunspot parameters, such as sunspot areas, number, etc, exhibit periodic changes with the solar cycle \citep{2015LRSP...12....4H}. Therefore, we also analyze the behaviour of $\rm{R}$ with time. In particular, we look for changes in $\rm{R}$ from cycle to cycle, as well as during the activity maxima and minima. We first analyze the distribution of $\rm{R}$ during solar activity maxima and minima. For better statistics, we merged the data from all cycles and from all the observatories. We define cycle maxima and minima as 12-month periods centered at the times of the maxima and minima in the 13-months running mean of the sunspot area (in the merged series), respectively. Now, there are significantly more bigger spots and groups during cycle maxima than during minima, while $\rm{R}$ depends on the group area (Figure~\ref{fig_with_tracking}b). Hence, to eliminate this size dependency, we first fit a lognormal function (\( \rm{R}\sim \exp{-\big[ \rm{(\log(A_{group}}))^2}/{2\sigma^2}\big] \)) to the data in Fig.~\ref{fig_with_tracking}b and then divide each measured $\rm{R}$ by the value this fit returns for the corresponding $\rm{A_{group}}$ to obtain the normalized $\rm{R}$ value. Figure~\ref{fig_cycle}a shows these normalized $\rm{R}$ distributions for cycle maxima and minima in blue and red, respectively. 
These distributions are nearly identical in shape and their mean, median and mode values (listed in the legend) are very similar. We further test the hypothesis that the two samples come from the same distribution by applying the two-sided Kolmogorov-Smirnov (K--S) test which checks for the equivalence of two datasets by comparing their empirical cumulative distribution functions (ECDFs) \citep{ber2014}. The test statistics, $\rm{D}$, is a measure of the difference between the ECDFs. The null hypothesis, that both samples come from the same underlying population, is rejected if $\rm{D}>D_{crit}$. On our two $\rm{R}$ distributions, $\rm{D}$=0.06 and D$_{crit}$=0.12, with p value being 0.44. Thus, in our case the null hypothesis is true, and we therefore conclude that no substantial variation in $\rm{R}$ is observed between the minimum and maximum of a sunspot cycle.

Finally, in Figure~\ref{fig_cycle}b, we show the cycle-to-cycle variation of the normalized $\rm{R}$. Filled circles represent the median values (second quartile) of $\rm{R}$ over each cycle, whereas the third and first quartiles of the corresponding $\rm{R}$ distributions are shown as the upper and lower limits of the error bars. We see no significant variation in $\rm{R}$ from cycle to cycle.

\section{Conclusion}

Analysis and understanding of the size distribution of sunspots and sunspot groups provides insights into the origin and evolution of solar magnetism. In this work, we have investigated the distribution of the spot sizes within sunspot groups. Historical sunspot archives from Kislovodsk (1954--2019), Pulkovo (1932--1991) and Debrecen (1974--2019) observatories have been analyzed to derive the ratio, $\rm{R}$, between the area of the biggest spot in a group ($\rm{A_{big\_spot}}$) and the total area of that group ($\rm{A_{group}}$). Our conclusions are:

\begin{enumerate}

\item
When considering all sunspot groups independently of their evolution, the distribution of $\rm{R}$ shows a clear peak around 0.5, i.e. a single big spot within a group occupies roughly 50\% of the whole group area (Fig~\ref{fig_non_tracking}). The distribution, however, shows a secondary peak at, or close to, unity, which we attribute to the group's evolution. Namely, the following spots in a group typically emerge later and decay faster, so that
during the initial and final stages of the group evolution, the leading spot is often the only observable spot in the group or there are only small following spots.

\item
By tracking individual groups over their lifetime we found that at the time when the group area reaches its maximum, the biggest spot in a group typically occupies about 60\% of the group area.
For half of all groups, $\rm{R}$ lies in the range between roughly 50\% and 70\% (Fig~\ref{fig_with_tracking}).

\item
We also find $\rm{R}$ to change with the group area, $\rm{A_{group}}$.
In smaller groups (30$\leq\rm{A_{group}}<$200~$\mu$Hem), $\rm{R}$ increases from 0.55 to 0.65. It then decreases slightly before settling at a nearly constant value of about $\approxeq$0.6 for $\rm{A_{group}}\geq$700~$\mu$Hem (Fig~\ref{fig_with_tracking}).

\item
  The number of individual umbrae within a given group, $\rm{N}$, increases with the group size as a power law function of the form $\rm{N}\propto\rm{A_{group}}^{0.58}$ (Fig~\ref{fig_ratio_number}). Since some spots have multiple umbrae, the number of individual spots within groups will increase somewhat weakly with the group size than the rate given by this function.

\item
 Extrapolation of our results to significantly bigger sunspot groups (such as unresolved starspot groups observed on highly active stars) suggests that a) such groups are composed of multiple spots, in agreement with the study by \citet{2004MNRAS.348..307S}; b) a significant fraction of 55-75\% of the group area is occupied by a single giant spot. Note that both these conclusions are, however, limited to spots formed by magnetic flux emerging in bipolar regions \citep[e.g.,][]{1996A&A...314..503S} and do not apply to starspots formed by magnetic flux being carried together, e.g. to the poles \citep{2001ApJ...551.1099S,2018A&A...620A.177I}. 

\item
The distribution of $\rm{R}$ does not change between solar activity maxima and minima, and we see no cycle to cycle variation (Fig~\ref{fig_cycle}).

\end{enumerate}

In summary, this paper deals with the distribution of spots within a sunspot group, providing constraints on the flux emergence processes. 
In Babcock-Leighton dynamo models, flux emergence plays a key role as it is responsible for both toroidal magnetic flux through the photosphere \citep{2020A&A...636A...7C} and for the generation of new poloidal magnetic flux through Joy’s law \citep{1919ApJ....49..153H}. Despite its critical role, the details of the emergence process remain poorly understood. For example, the cause of Joy’s law is not yet known: the Coriolis force is clearly implicated, but it is uncertain whether the underlying flows are those of the turbulent convective background \citep[e.g.,][]{1955ApJ...121..491P} or those internal to the rising flux tube \citep[e.g.,][]{1995ApJ...441..886C}. 
 We are now reaching the point where these possibilities can be modelled in detail. For example, numerical simulations starting from a magnetic flux concentration near the base of the convection zone and ending  with after the flux has emerged through the photosphere  into the solar atmosphere exist \citep{10.1093/mnras/staa844}, as do models where the magnetic and velocity fields from global dynamo simulations are coupled to models covering the emergence through the photosphere \citep{2017ApJ...846..149C}. Our findings of universal properties of this process, suggestive of a scale invariant emergence pattern, are an observational constraint on flux emergence process which will be useful in evaluating the different model simulations, and is thus  critical for understanding the solar dynamo. 
Our findings require that in order to match the observations the ratio of the area of the largest spot in a group relative to the area of all spots in the group should be approximately 0.55-0.65 for the dominant majority of sunspot groups. More conservatively, we can say that a very strong constraint emerging from our work is that the largest sunspot covers more than half the area of the whole sunspot group and hence also carries more than half of the magnetic flux. In fact, flux emergence is also a critical driver for the dynamics of the solar atmosphere, and our results also relevant for understanding and modelling the atmospheric response to flux emergence where the  distribution of the field at the surface is important. Further studies using a combination of simultaneous white-light and magnetic data (e.g. from SoHO/MDI \citep{sherrer1995}, SDO/HMI \citep{2012SoPh..275...17L} or in the future from SO/PHI \citep{2020A&A...642A..11S}) will additionally help us to understand the relation between the magnetic flux distribution and the observed spot distribution within a group. 

\section{Acknowledgement}
We thank the anonymous reviewer for the encouraging comments and helpful suggestions. We also thank the teams of the archives used in this study for all the work they had invested into obtaining and making these data available to the community.
 \bibliographystyle{aa}
 \bibliography{references_sunspot_group_distribution}
\end{document}